\documentclass[11pt,a4paper, twocolumn]{article}
\usepackage[utf8]{inputenc}
\usepackage[margin=.8in]{geometry} % For margin reduction
\usepackage[small]{caption}

\usepackage{amsmath}
\usepackage{graphicx}
\usepackage{chemfig}
\usepackage{authblk}
\usepackage[colorlinks=true, allcolors=blue]{hyperref}
\DeclareCaptionType[fileext=ext, name=Scheme, placement=tbph!]{scheme}
\author[1,2,*]{Laura Galleni}
\affil[1]
{Department of Chemistry, KU Leuven, Celestijnenlaan 200F, 3001 Leuven, Belgium}
\affil[2]
{Imec, Kapeldreef 75, 3001 Leuven, Belgium}

\author[1,2]{Faegheh S. Sajjadian}

\author[2]{Thierry Conard}

\author[1]{Daniel Escudero}

\author[2]{Geoffrey Pourtois}

\author[2,*]{Michiel J. van Setten}

\affil[*]{E-mail: laura.galleni@imec.be, michiel.vansetten@imec.be}

\title{Modeling X-ray Photoelectron Spectroscopy of Macromolecules Using \textit{GW}}
\date{}

\begin{document}
\twocolumn[
  \begin{@twocolumnfalse}
    \maketitle
    \begin{abstract}
  We propose a simple additive approach to simulate X-ray photoelectron spectra (XPS) of macromolecules based on the $GW$ method. Single-shot $GW$ ($G_0W_0$) is a promising technique to compute accurate core--electron binding energies (BEs). However, its application to large molecules is still unfeasible. To circumvent the computational cost of $G_0W_0$, we break the macromolecule into tractable building blocks, such as isolated monomers, and sum up the theoretical spectra of each component, weighted by their molar ratio. In this work, we provide a first proof of concept by applying the method to four test polymers and one copolymer, and show that it leads to an excellent agreement with experiments. The method could be used to retrieve the composition of unknown materials and study chemical reactions, by comparing the simulated spectra with experimental ones.
  \end{abstract}    
  \section*{\centering TOC Graphic}
    {\centering 
    \includegraphics[width=5cm]{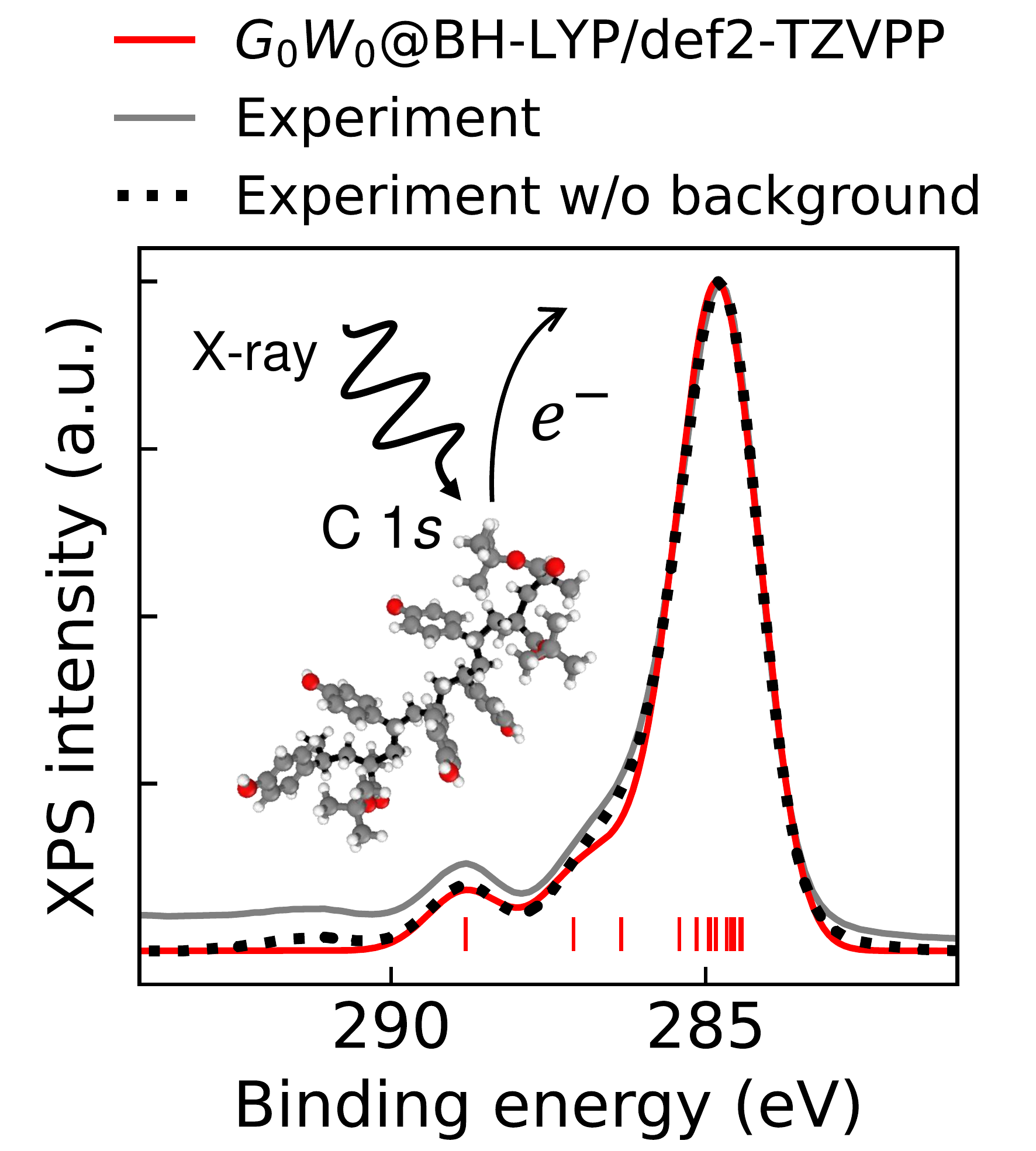}\par
    }
    \vspace{1cm}
 \end{@twocolumnfalse}
  ]

%%%%%%%%%%%%%%%%%%%%%%%%%%%%%%%%%%%%%%%%%%%%%%%%%%%%%%%%%%%%%%%%%%%%%
%% Start the main part of the manuscript here.
%%%%%%%%%%%%%%%%%%%%%%%%%%%%%%%%%%%%%%%%%%%%%%%%%%%%%%%%%%%%%%%%%%%%%

X-ray photoelectron spectroscopy (XPS) is a well-established experimental technique to study the chemical properties of solid, liquid, and gaseous materials \cite{bagus2013interpretationXPS, vanderheide2011xps}. In XPS, core-electron binding energies (BEs) are measured by means of the photoelectric effect.  
A typical XPS spectrum shows peaks at various BEs, with the peak height related to the number of electrons emitted from core orbitals at those BEs. The BEs are sensitive to the local chemical environment, such as the type of adjacent atoms, bonds, oxidation states. Therefore, absolute and relative peak positions can be used to retrieve information on the chemical composition of the material. The standard procedure to interpret XPS spectra consists in fitting the peaks to an envelope of Voigt functions with varying amounts of Gaussian and Lorentzian character \cite{major2020practicalXPSfitting}. However, the fitting procedure becomes increasingly challenging when more complicated compounds are considered, as the peaks tend to overlap and become indistinguishable due to the limited experimental resolution. The development of theoretical techniques to model XPS spectra is therefore important to support the interpretation of experiments  \cite{Vines2018Prediction, Norman2018Simulating, Besley2020DFTmethodsXspectr}. 

A common method to simulate XPS spectra is the Delta self-consistent field method ($\Delta$SCF) \cite{Norman2018Simulating, Vines2018Prediction}. In this technique, the BEs are calculated as the energy difference between the neutral species and the ionized species, where the energy of the latter is obtained after fixing a hole in a core orbital and relaxing the outer orbitals. Most of the $\Delta$SCF schemes make use of Hartree-Fock ($\Delta$HF) or density functional theory ($\Delta$DFT) \cite{Norman2018Simulating, Vines2018Prediction}. Similar techniques based on coupled-cluster calculations ($\Delta$CC) are also used on small molecules \cite{Zheng2019PerformanceDeltaCC}. The disadvantage of these approaches is the difficulty of artificially creating a hole in each core orbital, which can easily lead to convergence issues and become cumbersome for larger molecules. 

Another promising technique for calculating core-electron binding energy is $GW$, which is based on many body perturbation theory \cite{Aryasetiawan1998TheGWMethod, Onida2002, Golze2019Compendium, vanSetten2013gwImplementation}. $GW$ allows to calculate BEs directly from the one-particle Green's function, without explicitly generating holes in the core orbitals. The one-particle Green's function \(G(x t, x' t')\) is defined as follows: if $t>t'$, $G$ is the probability amplitude to find an electron at $x$ at time $t$ after addition of an electron at $x'$ at time $t'$, where $x$ and $x'$ include both spatial and spin coordinates; if $t<t'$, $G$ is the probability amplitude to find a hole at $x'$ at time $t'$ after removal of an electron at $x$ at time $t$. By definition, the poles of the Green's function correspond to the electron removal (attachment) energies as measured by (inverse) photoelectron spectroscopy. In principle, calculating the poles of the Green's function requires a fully self-consistent iterative solution of the Hedin equations \cite{hedin1965new}. To reduce the computational cost, the iterative procedure can be limited to a subset of variables, yielding different flavours of $GW$, such as single-shot $GW$ ($G_0W_0$), ev$GW$, sc$GW$, ev$GW_0$, and sc$GW_0$ \cite{Golze2019Compendium}. In the following, we consider only the least expensive $G_0W_0$, which corresponds to the first iteration of Hedin equations. As a starting point for this single-shot perturbation calculation, we consider Kohn-Sham states and eigenvalues calculated using DFT.

For valence electrons, the BEs calculated from $G_0W_0$ are in good agreement with experiments \cite{VanSetten2015GW100, Knight2016AccurateIpEaGW}. Only recently, the accuracy of the $G_0W_0$ method has been confirmed also for core-electron BEs \cite{VanSetten2018AssessingGWcore, Golze2018CoreBEfromGW, Golze2020Accurate}. 
Although $G_0W_0$ can potentially overcome the limitations of $\Delta$SCF and $\Delta$CC, its application to macromolecules is still unfeasible due to the high computational cost associated with these calculations. Therefore, approximations should be devised to use this method for larger molecular systems.
Here, we propose a simple additive approach to extend the range of applicability of $G_0W_0$ to macromolecules. As a proof of concept, we test the methodology on the four non-conjugated polymers in Scheme~\ref{sch:polymers}. These polymers have important applications in microelectronics as photoresists for extreme ultra-violet lithography \cite{manouras2020EUV}, besides being good examples of C, H, O materials containing carboxyl, methyl, hydroxyl, and benzene groups.

\begin{scheme}
    \centering
  \includegraphics[width=3in]{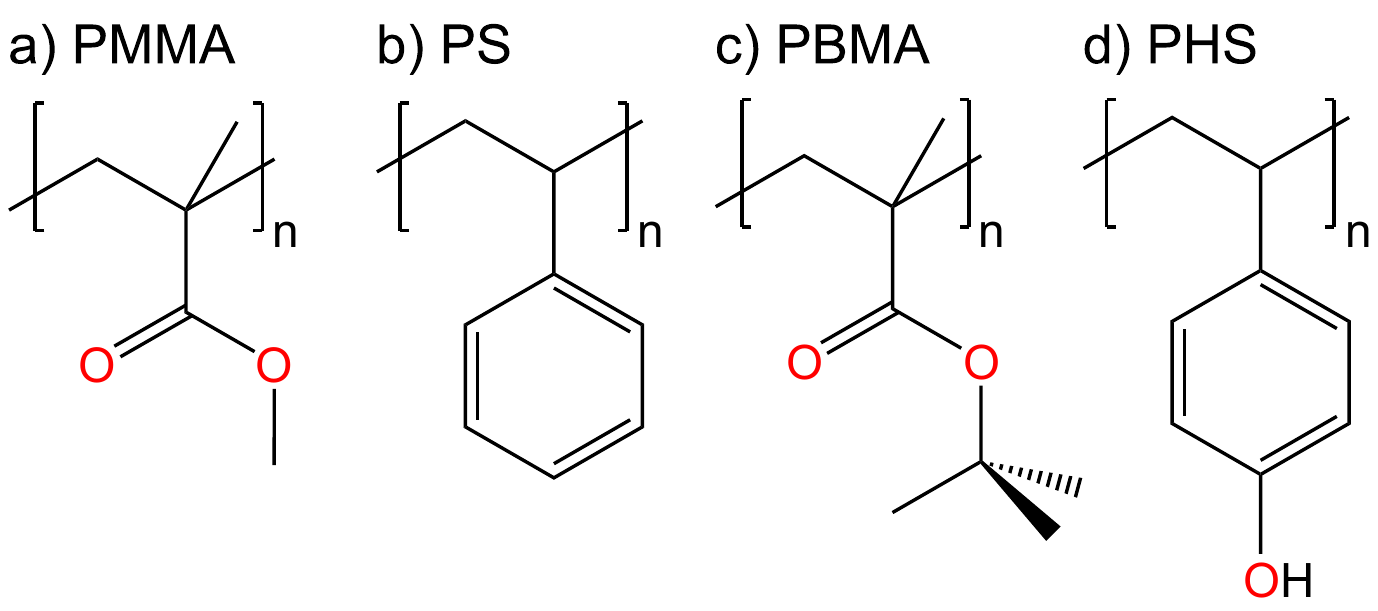}
  \caption{Polymers considered in this work.}
  \label{sch:polymers}
\end{scheme}

In first approximation, we can simulate a XPS spectrum as the sum of Gaussian peaks of width $\sigma$ centered at the BEs obtained from $G_0W_0$ calculations performed on top of DFT ($G_0W_0$@DFT). In reality, additional effects such as electron scattering in the material or electron-phonon interactions can add features to the experimental spectra, such as a broad background and satellite peaks, which are neglected here. 
\begin{scheme*}[]
    \centering
  \includegraphics[width=\textwidth]{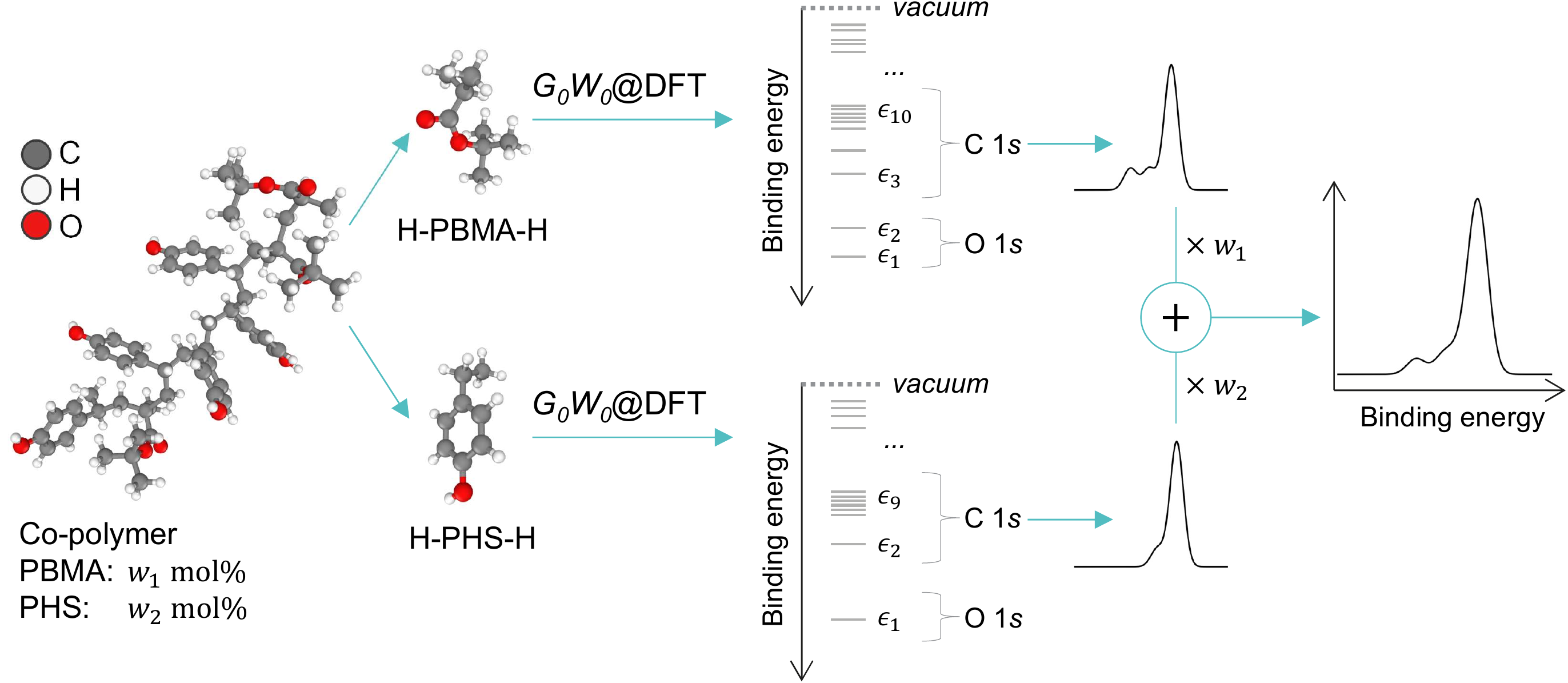}
  \caption{Schematic representation of the proposed methodology applied to hydrogen-terminated monomers of a poly[(t-butyl methacrylate)-\textit{co}-(p-hydroxystyrene)] copolymer (PBMA--PHS).}
  \label{sch:process}
\end{scheme*}
We assume that the dipole matrix elements are constant, or, equivalently, that all core electrons within the BE region of interest have the same photoionization cross-section \cite{Onida2002, Golze2019Compendium}. This is justified by the fact that all core orbitals under study are expected to show an approximately equivalent coupling with the X-ray photons, due to their $1s$-like spherical symmetry. Each orbital thus contributes to the photocurrent in proportion to the number of electrons occupying that orbital.
Therefore, the peaks are weighted by the spin degeneracy of the corresponding orbital, which corresponds to a factor $2$ for the molecules under study, as all orbitals are doubly-occupied. Since the wavefunctions of core electrons are strongly localized near the atomic nuclei, we can neglect the interaction between core electrons on distant atoms. This allows us to approximate the macromolecule as a collection of $N_{mol}$ independent building blocks of molar ratio $w_i$. The sum over all core-electron BEs can then be split over separate parts of the macromolecule, and the theoretical photocurrent simplifies to:
\begin{equation}
    I(E)= \mathcal{C} \sum_i^{N_{mol}} w_i \sum_{n_i}^{N_i} g_{n_i} e^{-\frac{(E-\epsilon_{n_i})^2}{2\sigma^2}}
    \label{eq:final_photocurrent}
\end{equation}
where $\mathcal{C}$ is a normalization factor to match the peak height with the experimental data, the first sum runs over the $N_{mol}$ building blocks of the macromolecule, and the second sum runs over the $N_i$ core orbitals of the $i$-th building block characterized by binding energy $\epsilon_{n_i}$ and spin degeneracy $g_{n_i}$. In this approximation, the calculation of the BEs can be performed separately on each building block, saving computational time. Of course, the building blocks should be chosen of the appropriate size in order to include the local chemistry while remaining tractable. In the case of polymers, this can be usually done by considering isolated monomers. A schematic of the calculation process is depicted in Scheme \ref{sch:process}.

Eq~\eqref{eq:final_photocurrent} is based on the assumption that the interactions between the building blocks are negligible. 
This assumption is expected to be generally valid for systems consisting of molecules that are not bonded together, and for non-conjugated polymeric materials where the core-electron BEs are not significantly affected by the monomers being connected to the polymer backbone. For conjugated polymers, however, this approximation could fail due to the strong electron delocalization. On top of this, the spectra of conjugated polymers are also complicated by additional effects such as oxidation and polaron states formation \cite{malitesta1995new}, which are currently not included in our approach.

For the four polymers under study, we investigate the validity of the independent-blocks assumption by computing the spectra of DFT optimized geometries of increasingly long polymer chains: monomers (H-$M$-H), dimers (H-$M_2$-H), and, where the computational power allows for it, trimers (H-$M_3$-H), where the two next atoms in the polymer backbone are replaced by hydrogen atoms. To assess the impact of adjacent backbone atoms, we also consider methyl-terminated monomers (CH$_3$-$M$-CH$_3$) and dimers (CH$_3$-$M_2$-CH$_3$), where the adjacent backbone atoms in the polymer chain are replaced with methyl groups. In the following, we will refer to these five types of cutting as “backbone corrections”. Of course, the addition of two methyl groups in CH$_3$-$M$-CH$_3$ and CH$_3$-$M_2$-CH$_3$ introduces two extra core levels in the calculated C $1s$ spectra. To avoid this artifact, we used a visualization software to identify the two C $1s$ orbitals localized on the two methyl groups, and removed the contributions of the corresponding BEs before computing the spectra. The core-electron Kohn-Sham orbitals and calculated BEs for the isolated monomers are depicted in Figures S1-S4.
\begin{figure*}
    \centering
    \includegraphics[width=\textwidth]{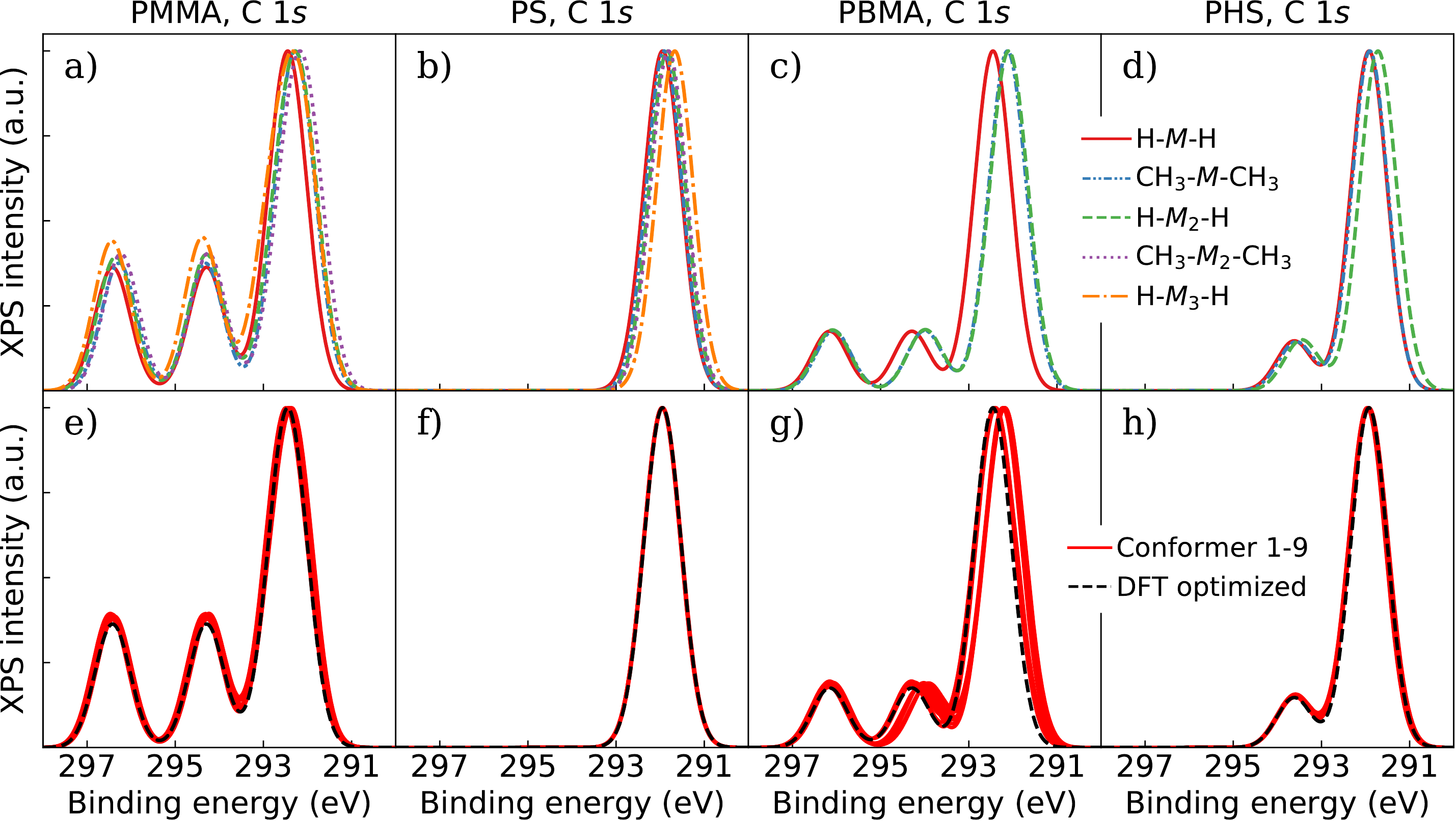}
    \caption{Theoretical C $1s$ XPS spectra of (a,e) PMMA, (b,f) PS, (c,g) PBMA, and (d,h) PHS calculated with $G_0W_0$@BH-LYP/def2-SVP and $\sigma=0.4$ eV. (a--d) Comparison between different backbone corrections: isolated monomers (H-$M$-H), dimers (H-$M_2$-H), trimers (H-$M_3$-H) and methyl-terminated monomers (CH$_3$-$M$-CH$_3$) and dimers (CH$_3$-$M_2$-CH$_3$). (e--h) Comparison between nine conformers of H-$M$-H and the optimized geometry obtained from a DFT relaxation in vacuum.}
    \label{fig:theory}
\end{figure*}

Figures~\ref{fig:theory}(a--d) show different backbone corrections for the four polymers under study. The theoretical XPS spectra are qualitatively similar regardless of the backbone correction. Quantitatively, different approximations lead to differences in peak positions of up to $0.4$ eV (Figure S5). Interestingly, the calculated peaks for styrene-based polymers, namely PS and PHS, tend to systematically move to smaller BEs when longer chains are considered. On the contrary, the effect of different backbone corrections on acrylate-based polymers, i.e. PMMA and PBMA, does not seem to follow a clear trend.

Another factor that might affect the calculated BEs is the molecular geometry. The presence of conformers in a sample could, in principle, be one of the causes of experimental broadening. To assess the impact of the molecular geometry on the core-electron BEs, we compare the spectra computed on nine different conformers of isolated monomers of the polymers under study. The conformers were extracted from a model polymer matrix of $50$ repeating units optimized with DFT (see Computational details). Hydrogen atoms were added to the extracted monomers along the backbone direction and their position was optimized with DFT while keeping the other atoms fixed. $G_0W_0$ was then used to calculate the BEs. Figures~\ref{fig:theory}(e--h) show that the peak variation over different conformers is smaller for styrene-based than for acrylate-based polymers, probably due to the rigidity of the aromatic group, which leads to a smaller conformational space. Overall, the variation due to the presence of conformers is generally below $0.04$ eV, except for PBMA, where it is up to $0.16$ eV (Tables S1-S4 and Figures S6-S9). These values are smaller than a typical experimental broadening ($0.5$ eV). Therefore, the peak broadening in XPS spectra most likely arises from other sources, such as finite excitation lifetime and limited experimental resolution.

The results shown above suggest that the XPS spectra of the polymers under study can be conveniently simulated by performing the calculations on much shorter chains. To assess the validity of this approximation, we compare our results with reported experimental spectra of two polymers: one acrylate-based (PMMA \cite{Louette2005}) and one styrene-based (PHS \cite{Chilkoti1990}) polymer. Both polymers were chosen for the existence of multiple features in the spectra and for the relatively small size of their monomers, which allow for a more complete assessment of various backbone corrections. For PMMA, both C $1s$ and O $1s$ edges were considered, whereas only the C $1s$ edge was investigated for PHS, as the O $1s$ edge shows only one peak and is thus less informative. 

A fitting procedure was performed to compare the calculated spectra with experiments. A rigid shift $\Delta$ was introduced to match the theoretical BEs with the experiment. The origin of the shift will be discussed below. The shift and the Gaussian broadening $\sigma$ were then fitted to minimize the differential area fraction $\mathcal{A}$ between the theoretical spectrum ($I$) calculated with Eq.~\eqref{eq:final_photocurrent} and the experimental data ($I_{exp}$) after subtraction of a Tougaard \cite{engelhard2020introductoryXPSbackground, tougaard1982influence} background ($I_{bg}$):

% \begin{equation}
%     \mathcal{A}(\Delta, \sigma)=\frac{1}{(E_2-E_1)\cdot I_{max}} \int_{E_1}^{E_2} \left|I_{exp}(E)-I_{bg}(E)-I(E-\Delta, \sigma)\right| \mathrm{d}E
%     \label{eq:differential_area}
% \end{equation}

\begin{equation}
\begin{split}
\mathcal{A}&(\Delta, \sigma)=\\
&\frac{1}{(E_2-E_1)\cdot I_{max}}\times \\
&\int_{E_1}^{E_2} \left|I_{exp}(E)-I_{bg}(E)-I(E-\Delta, \sigma)\right| \mathrm{d}E
\end{split}
\label{eq:differential_area}
\end{equation}

where $E_1$ and $E_2$ are chosen to contain all relevant features and $I_{max}=1$ as the spectra were normalized. 

\begin{figure*}
    \centering
    \includegraphics[width=\textwidth]{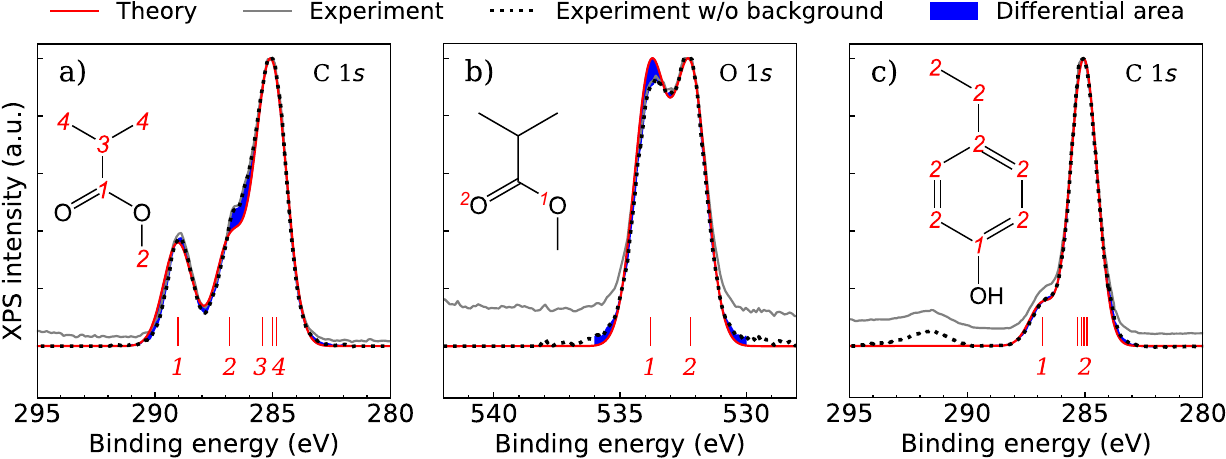}
    \caption{Theoretical spectra for (a) PMMA, C $1 s$, (b) PMMA, O $1 s$, and (c) PHS, C $1s$ fitted to experimental data from Ref.~ \cite{Louette2005} and  \cite{Chilkoti1990} after subtraction of a Tougaard \cite{engelhard2020introductoryXPSbackground, tougaard1982influence} background. The numbers in red identify sets of core-electron levels that are almost degenerate in energy and their location(s) on the molecule. Spectra were calculated on isolated monomers with $G_0W_0$@BH-LYP/def2-TZVPP. Fitted values: $\sigma=0.61$, $0.63$, and $0.55$ eV, $\Delta=6.13$, $5.58$, and $5.71$ eV, $\mathcal{A}=0.014$, $0.018$, and $0.012$, respectively; fit ranges: $291$--$282$, $536$--$530$, $289$--$282$ eV. The peak at $291.4$ eV in (c) is a $\pi\rightarrow\pi^*$ shakeup satellite  \cite{Chilkoti1990} and cannot be modeled with the technique presented here.}
    \label{fig:exp}
\end{figure*}

Figure~\ref{fig:exp} shows the fitted results on isolated monomers using $G_0W_0$@BH-LYP, revealing an excellent agreement with the experiment.
The comparison was performed for all possible combinations of three basis sets, eight hybrid functionals, and five backbone corrections (Figures S10-S15). Overall, the best results could be achieved by using hybrid functionals with approximately $50\%$ of exact exchange, such as BH-LYP ($50\%$), B2-PLYP ($53\%$), and M06-2X ($54\%$), similarly to the results of previous benchmarks on gas-phase molecules, where an optimal fraction of $45\%$ was found \cite{Golze2020Accurate}. Remarkably, the isolated monomer approximation is in general sufficient to reproduce the XPS profile of the full polymer. For the two polymers under study, improving on the basis set seems to be a better strategy than improving on the molecular cut, e.g. def2-TZVPP on isolated monomers yields a better agreement with the experiment than def2-SVP on dimers or trimers.

Before extending the methodology to copolymers, we briefly comment on the origin of the shift $\Delta$ between the theoretical and experimental BEs. First of all, we would like to emphasize that most of the chemical information in XPS spectra, i.e. the fingerprint of a material, is retrieved from \textit{relative}, not \textit{absolute}, binding energies. Therefore, the presence of a rigid shift, although undesirable, does not necessarily limit the applicability of the theoretical method. The results of this work show that $\Delta$ can be anywhere between $-2$ eV and $+8$ eV, depending on the polymer under study (PMMA or PHS), on the core orbital (C $1s$ or O $1s$), and on the level of theory, particularly the choice of basis sets and hybrid functionals for the underlying DFT calculations (Figures S10, S12 and S14). Although an absolute shift of $8$ eV is in general not negligible, in comparison with the absolute BE this corresponds to a relative error of only $1.5\%$ for O $1s$, and up to $2.8\%$ for C $1s$. 

The presence of a shift can be attributed to several factors on both theoretical and experimental sides. 
On the computational side, (i) the use of incomplete basis sets (def2-SVP, def2-TZVPP, and def2-QZVPP), and (ii) the amount of exact exchange in hybrid functionals ($10\%$ to $50\%$) can induce shifts of up to $2$ eV and $6$ eV, respectively. For solid materials, another source of discrepancy is (iii) the inconsistency between the zero energy references in theory and experiment, corresponding to the vacuum level and to the Fermi level, respectively. This inconsistency introduces a shift equal to the work function of the material. For non-conductive polymers as those investigated in this work, additional factors such as (iv) calibration issues due to the fact that the Fermi level is not visible in the spectra as it lies within the band gap, and (v) charging effects in the sample can contribute to shifts of several eVs.

In principle, (i) and (ii) can be tackled by extrapolation to the complete basis set limit and by using more expensive self-consistent $GW$ approaches. When only small gas-phase molecules are considered, solving (i) and (ii) can effectively reduce the overall shift to only a few tenths of eV, as reported by Golze et al. \cite{Golze2020Accurate}. However, when treating solids, the additional factors (iii-v) will overshadow any attempts to eliminate (i) and (ii).
In fact, referencing the BEs to HOMO energies would ideally solve (iii) and (iv), but this is not commonly done on the experimental side, also because of the decreased sensitivity of XPS in the valence region. Moreover, (v) cannot be easily eliminated. The calibration issues (iv) and (v) have been known for decades in the XPS community and have been historically tackled by shifting the experimental BEs to set the main C $1s$ peak at $285$ eV. This approach, although very practical, introduces in fact an additional unknown shift (vi), which is what ultimately makes it impossible to retrieve the absolute BEs from most published datasets. Therefore, for non-conductive polymers, only theoretical \textit{relative}, not \textit{absolute} BEs can be compared with experiments.

\begin{figure}[t]
    \centering
    \includegraphics[width=2.9in]{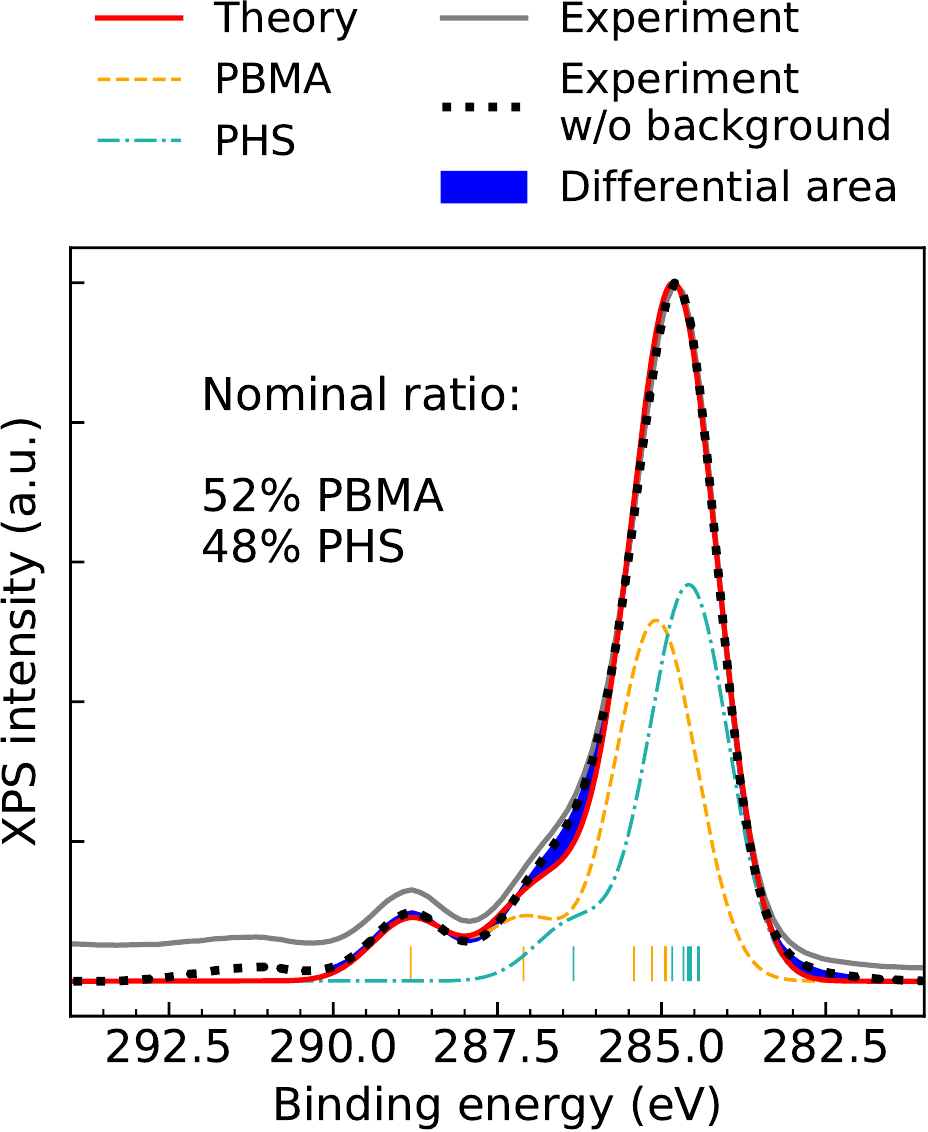}
    \caption{Theoretical spectra of a poly[(t-butyl methacrylate)-\textit{co}-(p-hydroxystyrene)] copolymer (PBMA--PHS) fitted to experimental data from this work after removal of a Tougaard background \cite{engelhard2020introductoryXPSbackground, tougaard1982influence}. The spectra were calculated on isolated monomers with $G_0W_0$@BH-LYP/def2-TZVPP and nominal monomer ratio. Fitted values: $\sigma=0.59$ eV, $\Delta=6.17$ eV, $\mathcal{A}=0.011$. The peak around $291$ eV  is a $\pi\rightarrow\pi^*$ shakeup satellite and cannot be modeled with the technique presented here.}
    \label{fig:fitratio}
\end{figure}

So far, we discussed the case of simple polymers. However, the same additive approach can, in principle, be applied also to copolymers, blends of polymers, as well as mixtures of polymers with non-bonded molecules. As explained above, in all these cases, we sum the spectra calculated for each isolated component, after multiplying the intensity by the molar ratio. To investigate the validity of our methodology for macromolecules consisting of multiple building blocks, we consider a poly[(t-butyl methacrylate)-\textit{co}-(p-hydroxystyrene)] copolymer (PBMA--PHS) with monomer ratios $w_{PBMA}=52\%$ and $w_{PHS}=48\%$. The fitting procedure was repeated using Eq.~\eqref{eq:differential_area} considering a Tougaard background model \cite{engelhard2020introductoryXPSbackground, tougaard1982influence}. Once again, the results are in good agreement with the experiment, as shown in Figure \ref{fig:fitratio}. The full benchmark results for different basis sets, hybrid functionals, and backbone corrections are reported in Figures S16 and S17.

Finally, we show that our methodology can also be used to extract the monomer ratio from the experimental spectra, with an error of about $10\%$. The fit was repeated by optimizing the parameter $w_{PBMA}$ (with $w_{PHS}=100\%-w_{PBMA}$) as well as $\Delta$ and $\sigma$. The results for different levels of theory are reported in Figure S18. Overall, hybrid functionals with around $50\%$ exact exchange yield fitted PBMA ratio within $10\%$ from the nominal ratio. However, the extracted ratio is very sensitive to the choice of the background model, resulting in a variation of about $10\%$ between linear \cite{engelhard2020introductoryXPSbackground}, Shirley \cite{engelhard2020introductoryXPSbackground, shirley1972high, proctor1982dataShirleyOriginal}, and Tougaard \cite{engelhard2020introductoryXPSbackground, tougaard1982influence} models (Table S5 and Figure S19). Further benchmark studies involving different copolymers and a range of monomer ratios are still required to assess the range of validity of this fitting procedure. Overall, our bottom-up approach can be a complementary method to standard fitting procedures to retrieve quantitative chemical composition from experimental spectra. 

In summary, we showed that accurate theoretical XPS spectra of macromolecules can be calculated using $G_0W_0$ by decomposing a macromolecule into tractable building blocks, such as isolated monomers, and summing up all contributions, weighted by their molar ratio. The calculated spectra are in excellent agreement with the experiments, provided that (i) hybrid functionals with around $50\%$ of exact exchange are used for the underlying DFT calculations, and (ii) a rigid shift is applied, to account for the inconsistent energy reference between theory and experiments.
Interestingly, the isolated monomers approximation is sufficient to reproduce XPS spectra of common acrylate- and styrene-based polymers and copolymers. The method may also work for other non-conjugated polymeric systems, peptides, mixtures of molecules, and other solid materials exhibiting weak inter-monomer and inter-molecular interactions. Due to the strong localization of core orbitals, the additive approach might in principle work also in combination with non-$GW$ methods, although we haven't tested this possibility. However, $GW$ is in general preferable to cumbersome methods such as $\Delta$SCF and $\Delta$CC, as $GW$ does not require the explicit creation of core holes. Overall, $G_0W_0$ seems to be the best approach in terms of accuracy and computational cost: to give an example of the efficiency of $G_0W_0$, consider that computing the BEs of the isolated monomers of the four polymers in Scheme~\ref{sch:polymers} with $G_0W_0$@BH-LYP  takes less than $3$ min with a def2-SVP basis set, and up to $50$ min with a def2-TZVPP basis set on a single core. Although not discussed in this work, our ongoing calculations show that this approach is also promising for mixtures of polymers with solvent-like molecules. Further studies are still needed to assess if the method can be extended to conjugated polymers, and how important screening effects are in extended systems, e.g. by comparing the proposed approach with electrostatic embedding methods. In the future, the strategy presented here could be applied to interpret XPS spectra of unknown materials and to study chemical reactions, by comparing the simulated spectra of various candidate compositions with the experimental one.

\section*{Computational details}
Molecular geometries were optimized with DFT using the PBE functional \cite{Perdew1996PBE} and a Gaussian basis set of triple-$\zeta$ valence quality (def2-TZVP) \cite{Weigend2005}. To optimize the model polymer structures of $50$ repeating units, we first run 200 steps of time-stamped force-bias Monte Carlo \cite{mees2012uniformTFMC} (TFMC) combined with DFT, then optimize using a SVP quality basis set, and finally optimize using a TZVP quality basis set. Binding energies were calculated with Turbomole 7.2 \cite{TURBOMOLE} using single-shot $GW$ ($G_0W_0$) \cite{vanSetten2013gwImplementation}, starting from Kohn--Sham orbitals and eigenvalues calculated by means of self-consistent field DFT. For benchmarking purposes, a combination of eight functionals and three basis sets were considered for the underlying DFT calculations, as implemented in Turbomole 7.2: TPSSH \cite{tao2003climbingTPSS, staroverov2003comparativeTPSSH}, B3-LYP \cite{becke1988densityb3-lyp1, lee1988developmentb3-lyp2, becke1993newB3-LYP}, PBE0 \cite{Perdew1996PBE, perdew1996rationalePBE0}, PBEH-3C \cite{grimme2015consistentPBEH-3C}, PW6B95 \cite{zhao2005designPW6B95}, BH-LYP \cite{becke1988densityb3-lyp1, lee1988developmentb3-lyp2, becke1993newBH-LYP}, B2-PLYP \cite{grimme2006semiempiricalB2PLYP}, M06-2X \cite{zhao2008constructionM06L}; Gaussian basis sets of split valence (def2-SVP), triple-$\zeta$ valence plus polarization (def2-TZVPP) and quadruple-$\zeta$ valence plus polarization (def2-QZVPP) quality combined with resolution of identity \cite{eichkorn1995auxiliary, Weigend2005, weigend2006accurate}. To fit theoretical spectra to the experimental data, Eq.~\eqref{eq:differential_area} was minimized by means of a modified Powell's algorithm, as implemented in the SCIPY package \cite{2020SciPy-NMeth}.
The linear, Shirley, and Tougaard backgrounds were computed with CasaXPS \cite{CASAXPSfairley2021systematic}.

\section*{Experimental details}
A $50$ nm thick film of poly[(t-butyl methacrylate)-\textit{co}-(p-hydroxystyrene)] copolymer ($52\%$ PBMA, $48\%$ PHS) was spin coated on a $5\times5$ cm$^2$ silicon wafer in a clean room environment followed by a post application bake at $90^\circ$C for $60$ s. 

The XPS measurements were performed in a VersaProbe III instrument from Ulvac-PHI using a monochromatized Al K$\alpha$ ($1486.6$ eV) photon beam. The sample was kept in the vacuum chamber for one night to ensure that all material outgassing was complete and that the vacuum level in the chamber was sufficiently low (in the range of $10^{-8}$ Pa) to carry out XPS.
To reduce the charging on the sample surface and minimize possible degradation due to X-rays, a large exposure spot size of $1000\times500$ $\mu m^2$ was used. The measurements were performed at a takeoff angle of $45^\circ$.

%%%%%%%%%%%%%%%%%%%%%%%%%%%%%%%%%%%%%%%%%%%%%%%%%%%%%%%%%%%%%%%%%%%%%
%% The "Acknowledgement" section can be given in all manuscript
%% classes.  This should be given within the "acknowledgement"
%% environment, which will make the correct section or running title.
%%%%%%%%%%%%%%%%%%%%%%%%%%%%%%%%%%%%%%%%%%%%%%%%%%%%%%%%%%%%%%%%%%%%%
\section*{Acknowledgements}
The authors thank Paul van der Heide (Imec) for fruitful discussions. The authors acknowledge funding from the Imec Industrial Affiliation Program (IIAP).

%%%%%%%%%%%%%%%%%%%%%%%%%%%%%%%%%%%%%%%%%%%%%%%%%%%%%%%%%%%%%%%%%%%%%
%% The same is true for Supporting Information, which should use the
%% suppinfo environment.
%%%%%%%%%%%%%%%%%%%%%%%%%%%%%%%%%%%%%%%%%%%%%%%%%%%%%%%%%%%%%%%%%%%%%
\section*{Supporting Information}
Core Kohn-Sham orbitals and corresponding BEs (Figures S1-S4); calculated XPS C $1s$ peak positions for different backbone corrections (Figure S5) and conformers (Tables S1-S4 and Figures S6-S9); fit results for all molecules and different levels of theory (Figures S10-S18); fitted monomer ratio for a PBMA--PHS copolymer after subtraction of different background models (Figure S19 and Table S5) (PDF)\\
XYZ structures of all molecules (TXT)\\
Calculated binding energies (JSON)

%%%%%%%%%%%%%%%%%%%%%%%%%%%%%%%%%%%%%%%%%%%%%%%%%%%%%%%%%%%%%%%%%%%%%
%% The appropriate \bibliography command should be placed here.
%% Notice that the class file automatically sets \bibliographystyle
%% and also names the section correctly.
%%%%%%%%%%%%%%%%%%%%%%%%%%%%%%%%%%%%%%%%%%%%%%%%%%%%%%%%%%%%%%%%%%%%%
\bibliographystyle{unsrt}
\small
\bibliography{bibliography_XPS}

\begin{thebibliography}{10}

\bibitem{bagus2013interpretationXPS}
Paul~S Bagus, Eugene~S Ilton, and Connie~J Nelin.
\newblock The interpretation of xps spectra: Insights into materials
  properties.
\newblock {\em Surf. Sci. Rep.}, 68(2):273--304, 2013.

\bibitem{vanderheide2011xps}
Paul Van~der Heide.
\newblock {\em X-Ray Photoelectron Spectroscopy: An Introduction to Principles
  and Practices}.
\newblock John Wiley \& Sons, Hoboken, 2011.

\bibitem{major2020practicalXPSfitting}
George~H Major, Neal Fairley, Peter~MA Sherwood, Matthew~R Linford, Jeff Terry,
  Vincent Fernandez, and Kateryna Artyushkova.
\newblock Practical guide for curve fitting in x-ray photoelectron
  spectroscopy.
\newblock {\em J. Vac. Sci. Technol. A}, 38(6):061203, 2020.

\bibitem{Vines2018Prediction}
Francesc Vi{\~{n}}es, Carmen Sousa, and Francesc Illas.
\newblock {On the Prediction of Core Level Binding Energies in Molecules,
  Surfaces and Solids}.
\newblock {\em Phys. Chem. Chem. Phys.}, 20(13):8403--8410, 2018.

\bibitem{Norman2018Simulating}
Patrick Norman and Andreas Dreuw.
\newblock {Simulating X-Ray Spectroscopies and Calculating Core-Excited States
  of Molecules}.
\newblock {\em Chem. Rev.}, 118(15):7208--7248, aug 2018.

\bibitem{Besley2020DFTmethodsXspectr}
Nicholas~A. Besley.
\newblock {Density Functional Theory Based Methods for the Calculation of X-Ray
  Spectroscopy}.
\newblock {\em Accounts Chem. Res.}, 53(7):1306--1315, jul 2020.

\bibitem{Zheng2019PerformanceDeltaCC}
Xuechen Zheng and Lan Cheng.
\newblock Performance of delta-coupled-cluster methods for calculations of
  core-ionization energies of first-row elements.
\newblock {\em J. Chem. Theory Comput.}, 15(9):4945--4955, 2019.

\bibitem{Aryasetiawan1998TheGWMethod}
F.~Aryasetiawan and O.~Gunnarsson.
\newblock {The GW Method}.
\newblock {\em Rep. Prog. Phys.}, 61(3):237--312, mar 1998.

\bibitem{Onida2002}
Giovanni Onida, Lucia Reining, and Angel Rubio.
\newblock {Electronic Excitations: Density-Functional Versus Many-Body
  Green’s-Function Approaches}.
\newblock {\em Rev. Mod. Phys.}, 74(2):601--659, jun 2002.

\bibitem{Golze2019Compendium}
Dorothea Golze, Marc Dvorak, and Patrick Rinke.
\newblock The gw compendium: A practical guide to theoretical photoemission
  spectroscopy.
\newblock {\em Front. Chem.}, 7:377, 2019.

\bibitem{vanSetten2013gwImplementation}
Michiel~J van Setten, Florian Weigend, and Ferdinand Evers.
\newblock The gw-method for quantum chemistry applications: Theory and
  implementation.
\newblock {\em J. Chem. Theory Comput.}, 9(1):232--246, 2013.

\bibitem{hedin1965new}
Lars Hedin.
\newblock New method for calculating the one-particle green’s function with
  application to the electron-gas problem.
\newblock {\em Phys. Rev.}, 139(3A):A796--A823, 1965.

\bibitem{VanSetten2015GW100}
Michiel~J. van Setten, Fabio Caruso, Sahar Sharifzadeh, Xinguo Ren, Matthias
  Scheffler, Fang Liu, Johannes Lischner, Lin Lin, Jack~R. Deslippe, Steven~G.
  Louie, Chao Yang, Florian Weigend, Jeffrey~B. Neaton, Ferdinand Evers, and
  Patrick Rinke.
\newblock {GW 100: Benchmarking $G_0W_0$ for Molecular Systems}.
\newblock {\em J. Chem. Theory Comput.}, 11(12):5665--5687, dec 2015.

\bibitem{Knight2016AccurateIpEaGW}
Joseph~W. Knight, Xiaopeng Wang, Lukas Gallandi, Olga Dolgounitcheva, Xinguo
  Ren, J.~Vincent Ortiz, Patrick Rinke, Thomas K{\"{o}}rzd{\"{o}}rfer, and Noa
  Marom.
\newblock {Accurate Ionization Potentials and Electron Affinities of Acceptor
  Molecules III: A Benchmark of GW Methods}.
\newblock {\em J. Chem. Theory Comput.}, 12(2):615--626, feb 2016.

\bibitem{VanSetten2018AssessingGWcore}
Michiel~J. van Setten, Ramon Costa, Francesc Vi{\~{n}}es, and Francesc Illas.
\newblock {Assessing GW Approaches for Predicting Core Level Binding Energies}.
\newblock {\em J. Chem. Theory Comput.}, 14(2):877--883, feb 2018.

\bibitem{Golze2018CoreBEfromGW}
Dorothea Golze, Jan Wilhelm, Michiel~J. van Setten, and Patrick Rinke.
\newblock {Core-Level Binding Energies From GW: An Efficient Full-Frequency
  Approach Within a Localized Basis}.
\newblock {\em J. Chem. Theory Comput.}, 14(9):4856--4869, sep 2018.

\bibitem{Golze2020Accurate}
Dorothea Golze, Levi Keller, and Patrick Rinke.
\newblock {Accurate Absolute and Relative Core-Level Binding Energies From GW}.
\newblock {\em J. Phys. Chem. Lett.}, 11(5):1840--1847, 2020.

\bibitem{manouras2020EUV}
Theodore Manouras and Panagiotis Argitis.
\newblock High sensitivity resists for euv lithography: A review of material
  design strategies and performance results.
\newblock {\em Nanomaterials}, 10(8):1593, 2020.

\bibitem{malitesta1995new}
C~Malitesta, I~Losito, L~Sabbatini, and PG~Zambonin.
\newblock New findings on polypyrrole chemical structure by xps coupled to
  chemical derivatization labelling.
\newblock {\em J. Electron Spectrosc. Relat. Phenom.}, 76:629--634, 1995.

\bibitem{Louette2005}
Pierre Louette, Frederic Bodino, and Jean-Jacques Pireaux.
\newblock {Poly(methyl Methacrylate) (PMMA) XPS Reference Core Level and Energy
  Loss Spectra}.
\newblock {\em Surf. Sci. Spectra}, 12(1):69--73, 2005.

\bibitem{Chilkoti1990}
Ashutosh Chilkoti, David~G. Castner, Buddy~D. Ratner, and David Briggs.
\newblock {Surface Characterization of a Poly(styrene/P-Hydroxystyrene)
  Copolymer Series Using X-Ray Photoelectron Spectroscopy, Static Secondary Ion
  Mass Spectrometry, and Chemical Derivatization Techniques}.
\newblock {\em J. Vac. Sci. Technol. A}, 8(3):2274--2282, 1990.

\bibitem{engelhard2020introductoryXPSbackground}
Mark~H Engelhard, Donald~R Baer, Alberto Herrera-Gomez, and Peter~MA Sherwood.
\newblock Introductory guide to backgrounds in xps spectra and their impact on
  determining peak intensities.
\newblock {\em J. Vac. Sci. Technol. A}, 38(6):063203, 2020.

\bibitem{tougaard1982influence}
Sven Tougaard and Peter Sigmund.
\newblock Influence of elastic and inelastic scattering on energy spectra of
  electrons emitted from solids.
\newblock {\em Phys. Rev. B}, 25(7):4452--4466, 1982.

\bibitem{shirley1972high}
Dave~A Shirley.
\newblock High-resolution x-ray photoemission spectrum of the valence bands of
  gold.
\newblock {\em Phys. Rev. B}, 5(12):4709--4714, 1972.

\bibitem{proctor1982dataShirleyOriginal}
Andrew Proctor and Peter~MA Sherwood.
\newblock Data analysis techniques in x-ray photoelectron spectroscopy.
\newblock {\em Anal. Chem.}, 54(1):13--19, 1982.

\bibitem{Perdew1996PBE}
John~P. Perdew, Kieron Burke, and Matthias Ernzerhof.
\newblock {Generalized Gradient Approximation Made Simple}.
\newblock {\em Phys. Rev. Lett.}, 77(18):3865--3868, oct 1996.

\bibitem{Weigend2005}
Florian Weigend and Reinhart Ahlrichs.
\newblock {Balanced Basis Sets of Split Valence, Triple Zeta Valence and
  Quadruple Zeta Valence Quality for H to Rn: Design and Assessment of
  Accuracy}.
\newblock {\em Phys. Chem. Chem. Phys.}, 7(18):3297--3305, 2005.

\bibitem{mees2012uniformTFMC}
Maarten~J Mees, Geoffrey Pourtois, Erik~C Neyts, Barend~J Thijsse, and
  Andr{\'e} Stesmans.
\newblock Uniform-acceptance force-bias monte carlo method with time scale to
  study solid-state diffusion.
\newblock {\em Phys. Rev. B}, 85(13):134301, 2012.

\bibitem{TURBOMOLE}
{TURBOMOLE V7.2 2017}, a development of {University of Karlsruhe} and
  {Forschungszentrum Karlsruhe GmbH}, 1989-2007, {TURBOMOLE GmbH}, since 2007;
  available from {\tt http://www.turbomole.com}.

\bibitem{tao2003climbingTPSS}
Jianmin Tao, John~P Perdew, Viktor~N Staroverov, and Gustavo~E Scuseria.
\newblock Climbing the density functional ladder: Nonempirical
  meta-–generalized gradient approximation designed for molecules and solids.
\newblock {\em Phys. Rev. Lett.}, 91(14):146401, 2003.

\bibitem{staroverov2003comparativeTPSSH}
Viktor~N Staroverov, Gustavo~E Scuseria, Jianmin Tao, and John~P Perdew.
\newblock Comparative assessment of a new nonempirical density functional:
  Molecules and hydrogen-bonded complexes.
\newblock {\em J. Chem. Phys.}, 119(23):12129--12137, 2003.

\bibitem{becke1988densityb3-lyp1}
Axel~D Becke.
\newblock Density-functional exchange-energy approximation with correct
  asymptotic behavior.
\newblock {\em Phys. Rev. A}, 38(6):3098--3100, 1988.

\bibitem{lee1988developmentb3-lyp2}
Chengteh Lee, Weitao Yang, and Robert~G Parr.
\newblock Development of the colle-salvetti correlation-energy formula into a
  functional of the electron density.
\newblock {\em Phys. Rev. B}, 37(2):785--789, 1988.

\bibitem{becke1993newB3-LYP}
Axel~D Becke.
\newblock Density-functional thermochemistry. iii. the role of exact exchange.
\newblock {\em J. Chem. Phys.}, 98(7):5648--5652, 1993.

\bibitem{perdew1996rationalePBE0}
John~P Perdew, Matthias Ernzerhof, and Kieron Burke.
\newblock Rationale for mixing exact exchange with density functional
  approximations.
\newblock {\em J. Chem. Phys.}, 105(22):9982--9985, 1996.

\bibitem{grimme2015consistentPBEH-3C}
Stefan Grimme, Jan~Gerit Brandenburg, Christoph Bannwarth, and Andreas Hansen.
\newblock Consistent structures and interactions by density functional theory
  with small atomic orbital basis sets.
\newblock {\em J. Chem. Phys.}, 143(5):054107, 2015.

\bibitem{zhao2005designPW6B95}
Yan Zhao and Donald~G Truhlar.
\newblock Design of density functionals that are broadly accurate for
  thermochemistry, thermochemical kinetics, and nonbonded interactions.
\newblock {\em J. Phys. Chem. A}, 109(25):5656--5667, 2005.

\bibitem{becke1993newBH-LYP}
Axel~D Becke.
\newblock A new mixing of hartree–fock and local density-functional theories.
\newblock {\em J. Chem. Phys.}, 98(2):1372--1377, 1993.

\bibitem{grimme2006semiempiricalB2PLYP}
Stefan Grimme.
\newblock Semiempirical hybrid density functional with perturbative
  second-order correlation.
\newblock {\em J. Chem. Phys.}, 124(3):034108, 2006.

\bibitem{zhao2008constructionM06L}
Yan Zhao and Donald~G Truhlar.
\newblock Construction of a generalized gradient approximation by restoring the
  density-gradient expansion and enforcing a tight lieb–oxford bound.
\newblock {\em J. Chem. Phys.}, 128(18):184109, 2008.

\bibitem{eichkorn1995auxiliary}
Karin Eichkorn, Oliver Treutler, Holger {\"O}hm, Marco H{\"a}ser, and Reinhart
  Ahlrichs.
\newblock Auxiliary basis sets to approximate coulomb potentials.
\newblock {\em Chem. Phys. Lett.}, 240(4):283--290, 1995.

\bibitem{weigend2006accurate}
Florian Weigend.
\newblock Accurate coulomb-fitting basis sets for h to rn.
\newblock {\em Phys. Chem. Chem. Phys.}, 8(9):1057--1065, 2006.

\bibitem{2020SciPy-NMeth}
Pauli Virtanen, Ralf Gommers, Travis~E. Oliphant, Matt Haberland, Tyler Reddy,
  David Cournapeau, Evgeni Burovski, Pearu Peterson, Warren Weckesser, Jonathan
  Bright, St{\'e}fan~J. {van der Walt}, Matthew Brett, Joshua Wilson, K.~Jarrod
  Millman, Nikolay Mayorov, Andrew R.~J. Nelson, Eric Jones, Robert Kern, Eric
  Larson, C~J Carey, {\.I}lhan Polat, Yu~Feng, Eric~W. Moore, Jake
  {VanderPlas}, Denis Laxalde, Josef Perktold, Robert Cimrman, Ian Henriksen,
  E.~A. Quintero, Charles~R. Harris, Anne~M. Archibald, Ant{\^o}nio~H. Ribeiro,
  Fabian Pedregosa, Paul {van Mulbregt}, and {SciPy 1.0 Contributors}.
\newblock {{SciPy} 1.0: Fundamental Algorithms for Scientific Computing in
  Python}.
\newblock {\em Nat. Methods}, 17:261--272, 2020.

\bibitem{CASAXPSfairley2021systematic}
Neal Fairley, Vincent Fernandez, Mireille Richard-Plouet, Catherine
  Guillot-Deudon, John Walton, Emily Smith, Delphine Flahaut, Mark Greiner,
  Mark Biesinger, Sven Tougaard, et~al.
\newblock Systematic and collaborative approach to problem solving using x-ray
  photoelectron spectroscopy.
\newblock {\em Appl. Surf. Sci. Adv.}, 5:100112, 2021.

\end{thebibliography}

\end{document}